\documentclass[article]{aa}
\usepackage{graphicx}
\usepackage{longtable}
\newcommand{\lsun}{log$L/L_{\odot}\,$}
\newcommand{\msun}{$M_{\odot}\,$}

\begin{document}

   \title{Synthetic properties of bright metal-poor variables. II. BL Her stars.}

\author{M. Di Criscienzo \inst{1,2}, F. Caputo \inst{3}, M. Marconi \inst{1}, S. Cassisi \inst{4}}

\institute{\bf{INAF-}Osservatorio Astronomico di Capodimonte, Via Moiariello 16,
80131 Napoli, Italy;\\ \and Universit\`a degli Studi di Roma
``Tor Vergata'', Via della Ricerca Scientifica 1,
00133, Roma, Italy;  \\
\and \bf{INAF-}Osservatorio Astronomico di Roma, Via Frascati 33, 00040 Monte Porzio
Catone, Italy;\\
\and \bf{INAF-}Osservatorio Astronomico di Collurania, Via Maggini, 64100 Teramo, Italy;
 }

\date{}

   \abstract{We investigate the properties of the so-called BL Her stars,
i.e., Population II Cepheids with periods shorter than 8 days, using
updated pulsation models and evolutionary tracks computed adopting a
metal abundance in the range of $Z$=0.0001 to $Z$=0.004. We derive
the predicted Period-Magnitude ($PM$) and Period-Wesenheit ($PW$)
relations at the various photometric bands and we show that the
slopes of these relations are in good agreement with the slopes
determined by observed variables in Galactic globular clusters,
independently of the adopted $M_V$(RR)-[Fe/H] relation to get the
cluster RR Lyrae-based distance. Moreover, we show that also the
distances provided by the predicted $PM$ and $PW$ relations for BL
Her stars agree within the errors with the RR Lyrae based values.
The use of the predicted relations with W Vir stars, which are
Population II Cepheids with periods longer than 8 days, provides no
clear evidence for or against a change in the $PM$ and $PW$ slopes
around $P \sim$ 10 days. }


\authorrunning{Di Criscienzo \it{et al.}}

\titlerunning{Synthetic properties of bright metal-poor variables. II. }

\maketitle

%________________________________________________________________

\pagebreak
\section{Introduction}
Population II pulsating variables play a fundamental role in our
understanding of the properties of old stellar populations as well
as in the definition of the  cosmic distance scale. Among them, RR
Lyrae stars are definitively the most abundant  and the ones
currently used as tracers of dynamical and chemical  properties.
Moreover, they are used as standard candles to establish the
globular cluster distance scale and they provide the calibration of
secondary distance indicators such, e.g., the Globular Cluster
Luminosity Function in external galaxies (see Di Criscienzo et al.
2005 and references therein).\\
However, other classes of radial pulsators are actually observed in
globular clusters and similar metal-poor stellar fields. In the
current nomenclature, they are named Population II Cepheids (P2Cs)
and Anomalous Cepheids (ACs): the former ones, with periods $P$ from
$\sim$ 1 to $\sim$ 25 days, are observed in clusters with few RR
Lyrae stars and blue Horizontal Branch (HB) morphology, while the
latter, with $\sim 0.3 \le P\le \sim$ 2 days, are observed in the
majority of the Local Group dwarf galaxies which have been surveyed
for variable stars. These two classes are both brighter but either
less massive (P2Cs) or more massive (ACs) than RR Lyrae stars with
similar metal content.\\
In three previous papers (Marconi, Fiorentino \& Caputo 2004; Caputo
et al. 2004; Fiorentino et al. 2006) dealing with the investigation
of the ACs and their role as distance indicators, we discussed the
pulsation and evolution properties of these variables and we showed that they
originate from $Z \leq$ 0.0004 central He-burning models more
massive than $\sim$ 1.3\msun which evolve through the pulsation
region at luminosity and effective temperature which increase, on
average, as the mass increases. Given the basic equation for radial
pulsation ($P\rho^{1/2}=Q$, where $\rho$ is the star density and $Q$
is the pulsation constant), the effect of the higher luminosity is
balanced by the larger mass and temperature and consequently, in spite of
their bright luminosity, the ACs show periods that are not significantly longer than
those typical of RR Lyrae stars.\\
Concerning the P2Cs, which are often separated\footnote {In
their recent reviev, Sandage \& Tammann (2006) adopt a different
classification. However, in the present paper we will use the
classical one.} into BL Her stars (log$P<$ 1) and  W Vir stars
(log$P>$ 1), several authors (e.g., Gingold 1985; Bono, Caputo \&
Santolamazza 1997; Wallerstein \& Cox 1984; Harris 1985; Wallerstein
1990, 2002) have already suggested that they originate from hot,
low-mass stellar structures which started the main central
He-burning phase in the blue side of the RR Lyrae gap and now evolve
toward the Asymptotic Giant Branch crossing the pulsation region
with the luminosity and the effective temperature that increases and
decreases, respectively, with decreasing the mass: for this reason,
these bright low-mass pulsators should reach periods of several
days. Moreover, as also shown by Caputo et al. 2004 (see their Fig.
4) on theoretical ground, at fixed period the ACs are more luminous
than P2Cs, a feature which is at the origin of their supposed
``anomaly''.

On the observational side, Nemec, Nemec \& Lutz (1994) derived metal
dependent Period-Luminosity ($PL$) relations in various photometric
bands, suggesting that observed P2Cs pulsate either in the
fundamental and in the first-overtone mode and  that the slopes of
the $PL$ relations are significantly different for the two modes. On
the other hand, on the basis of a sample of P2Cs identified in the
OGLE-II variable star catalogue for the Galactic bulge fields,
Kubiack \& Udalsky (2003) found that all the observed stars, which
have periods from $\sim$ 0.7 to about 10 days,  follow the same $PL$
relation. Similar results are derived by Pritzl et al. (2003) and
Matsunaga et al. (2006) for P2Cs in Galactic globular clusters.
Furthermore, these last two investigations support the hypothesis
that the same $PL$ relation holds for BL Her and W Vir stars,
without a steepening of the slope for periods longer than $P\sim$ 10
days, as earlier suggested by McNamara (1995).

From the theoretical point of view, the pulsation models by Buchler
\& Moskalik (1992) and Buchler \& Buchler (1994), as based on a
linear and nonlinear radiative analysis, showed that the blue edge
for first-overtone pulsation was very close ($\le$ 100 K) to the
fundamental one, producing a very narrow region of FO-only
pulsation. More recently, Bono, Caputo \& Santolamazza (1997)
computed nonlinear convective models, finding a good agreement
between the predicted and the observed boundaries of the P2C
instability strip and suggesting that the observed variables are
pulsating in the fundamental mode with a typical mass of $\sim$
0.52-0.59\msun.\\
However, the Bono, Caputo \& Santolamazza (1997) nonlinear
convective models, although able to provide reliable information
also on the red edge of pulsation region, were limited to a quite
restricted range of stellar parameters and adopted an old input
physics (see Bono \& Stellingwerf 1994 for details). For this
reason, following our program dealing with a homogeneous study of
radially pulsating stars with various chemical composition, mass and
luminosity, in the present paper we discuss the results of updated
pulsation models with mass 0.50-0.65\msun and luminosity
\lsun=1.81-2.41 in order to build a sound theoretical scenario for
the analysis of the P2Cs. In particular, we will derive the
predicted relations connecting evolutionary and pulsation properties
for BL Her stars and we will verify their use as distance
indicators.

The paper is organized as follows: in Section 2, we present the
evolutionary and pulsation models, while in Section 3 we deal with
the evolution-pulsation connection and we give the predicted
relations. The comparison with observed variables is presented in
Section 4 and the conclusion close the paper.

\section{Theoretical framework}

\begin{table}
\caption{Basic parameters of the pulsation models and resulting
effective temperatures at the edges for fundamental and
first-overtone pulsation (see Note). If no value is given at FOBE
and FORE, we found only fundamental models. A helium abundance
$Y$=0.24 and a mixing-length parameter $l/H_p$=1.5 has been adopted.
Mass and luminosity are in solar units.}
% title of Table
\label{table:1}      % is used to refer this table in the text
\begin{center}                          % used for centering table
\begin{tabular}{lcccccc}        % centered columns (4 columns)
\hline\hline                 % inserts double horizontal lines
$Z$& $M$ & log$L$ & FOBE$^a$&FBE$^b$&FORE$^c$&FRE$^d$ \\    % table heading
\hline                        % inserts single horizontal line
0.0001& 0.60 & 1.95 &-&6850&-&5750\\
      &      & 2.05 &-&6750&-&5650\\
      &      & 2.15 &-&6750&-&5550\\
      & 0.65 & 1.91 &6950&6850&6050&5750\\
      &      & 2.01 &6750&6850&6250&5750\\
      &      & 2.11 &-&6750&-&5550\\
\hline
0.001 & 0.50 & 2.11 &-&6650&-&5450\\
    & 0.50 & 2.41 &-&6350&-&5150\\
    & 0.55 & 1.81 &6850&6850&6350&5650\\
      &      & 1.91 &-&6850&-&5550\\
      &      & 2.01 &-&6750&-&5450\\
      & 0.65 & 1.81 &7050&6750&6650&5750\\
      &      & 1.91 &6850&6750&6150&5650\\
      &      & 2.01 &6650&6850&6350&5650\\
\hline
0.004 & 0.55 & 1.81 &-&6950&-&5750 \\
      &      & 1.91 &-&6850&-&5650 \\
      &      & 2.01 &-&6750&-&5450 \\
\hline
\end{tabular}
\end{center}
Note: a) First Overtone Blue Edge; b) Fundamental Blue Edge; c)
First Overtone Red Edge; d) Fundamental Red Edge.
\end{table}

The pulsation models computed for the present paper, as listed in
Table 1, adopt the same nonlinear, nonlocal and time-dependent
convective hydrodynamical code and the same physical assumptions
(i.e., equation of state and opacity tables)  already used for the
analysis of Classical Cepheids (Caputo et al. 2005; Marconi, Musella
\& Fiorentino 2005; Fiorentino et al. 2007), RR Lyrae stars (Marconi
et al. 2003; Di Criscienzo et al. 2004) and Anomalous Cepheids
(Marconi, Fiorentino \& Caputo 2004; Fiorentino et al. 2006). In
those papers, several relations connecting pulsational and
evolutionary parameters were derived, whose slopes show a general
consistency with the observed values. Moreover, they gave also a
good agreement with the features of observed light curves of
Classical Cepheids (Bono, Castellani \& Marconi 2002) and RR Lyrae
stars (Bono, Castellani \& Marconi 2000; Castellani, Degl'Innocenti
\& Marconi 2002; Di Criscienzo, Marconi \& Caputo 2004; Marconi \&
Clementini 2005). On this ground, our pulsation models appear in
principle able to provide reliable information on the structural
parameters of observed variables and, in turn, on the distance
modulus, although we cannot exclude that the theoretical results are
affected by unknown systematic errors and that further work is need
to refine our knowledge. However, we wish to emphasize that our
computations provide a homogeneous pulsational scenario for the
study of complex stellar systems where a variety of pulsating stars
can be observed.

The model sequences discussed in the present paper are computed as
one parameter families with constant chemical composition, mass and
luminosity, by varying the effective temperature $T_e$ by steps of
100 K. These models, which adopt a value of the mixing length
parameter $l/H_p$=1.5 to close the system of convective and dynamic
equations, are fully described by Marconi \& Di Criscienzo (2007)
and here we report only the results relevant for the purpose of the
present paper. For the sake of the following discussion, let us
firstly make clear that increasing (decreasing) by 100 K the
effective temperature of the computed bluest (reddest) fundamental
(F) or first-overtone (FO) model yields non-pulsating structures in
the corresponding mode. Accordingly, we adopt the effective
temperature of the computed bluest FO and F model, increased by 50
K, as the first-overtone (FOBE) and the fundamental blue edge (FBE)
respectively, and the effective temperature of the reddest FO and F
model, decreased by 50 K, as the first-overtone (FORE) and the
fundamental red edge (FRE) respectively. This yields that the
effective temperatures given in Table 1 have the intrinsic
uncertainty of $\pm$ 50 K.

Starting with the models with 0.65\msun and log$L/L_{\odot}$=1.81,
we note that they follow the well known behaviour of RR Lyrae stars
in that FO models are generally bluer than the F ones, but with the
FORE redder than the FBE. As a consequence, we have that: {\it a)}
the limits of the whole pulsation region are described by the FOBE
and the FRE; {\it b)} both the pulsation modes are stable in the
middle zone delimited by the FBE and the FORE; {\it c)} F-only
pulsators are located between the FRE and the FORE and FO-only
pulsators between the FBE and the FOBE. By increasing the
luminosity, the whole pulsation region moves towards the red, but
with a significant shrinking of the FO-only pulsation region. Based on the
values listed in Table 1,
the difference FOBE$-$FBE is $\sim$ +300 K at \lsun=1.81 and $\sim$ +100 K at
\lsun=1.91. A further increase of the
luminosity yields that the FOBE becomes redder than the FBE
(with a difference FOBE$-$FBE $\sim -$150 K at \lsun=2.01), with
the total disappearance of stable FO models at \lsun$\ge$ 2.11.

Varying the mass, we note that with 0.60\msun no FO model is
stable at \lsun$\ge$ 1.95, while with 0.55\msun we get FOBE=FBE at
\lsun=1.80 and only F models above this luminosity level. On the
other hand, by relying on the computations discussed by Marconi et
al. (2003) and Di Criscienzo et al. (2004), we recall that for
models with 0.80\msun the difference FOBE$-$FBE is $\sim$ +400 K at
\lsun=1.72 and $\sim$ +200 K at \lsun=1.91.

In summary, the results listed in Table 1 confirm earlier
suggestions (see Tuggle \& Iben 1972; Bono, Castellani \&
Stellingwerf 1995; Bono, Caputo \& Santolamazza 1997) that for each
given mass and helium content there exists an ``intersection''
luminosity $L_{IP}$ where FOBE=FBE, and that above this luminosity
only the fundamental mode is stable. On this ground, one has that
the red limit of the instability strip is always determined by the
FRE, while the blue limit is given by the FOBE or the FBE depending
on whether the luminosity is fainter or brighter, respectively, than
$L_{IP}$.

Based on present computations and the quoted RR Lyrae models, we
estimate at $Y$=0.24
$$\log L_{IP}\sim 2.3+1.9\log M,\eqno(1)$$
\noindent
where mass and luminosity are in solar units.
Concerning the limits of the instability strip, we adopt
$$\log T_e(FOBE)=3.970(\pm0.004)-0.057\log L+0.094\log M,\eqno(2)$$
\noindent as determined by Marconi et al. (2003) from pulsation
models with $L<L_{IP}$, while a linear interpolation through the
present results gives
$$\log T_e(FBE)=3.912(\pm0.007)-0.035\log L+0.048\log M\eqno(3)$$
$$\log T_e(FRE)=3.925(\pm0.006)-0.075\log L+0.118\log M,\eqno(4)$$

\noindent where the uncertainties include the
intrinsic uncertainty of $\pm$ 50 K on
the FOBE, FBE and FRE temperatures.
Moreover, we derive that the pulsation equation for the
fundamental mode can be approximated as
$$\log P_F=11.579(\pm0.015)+0.89\log L-0.89\log M-3.54\log T_e\eqno(5)$$
\noindent while for the few first-overtone models we get
log$P_{FO}\sim$ log$P_F-$0.12, at constant mass, luminosity and
effective temperature.

Before proceeding, it is worth mentioning that the adopted
helium content $Y$=0.24 accounts for the most recent estimate
(Cassisi et al. 2003; Salaris et al. 2004) based on measurements of
the $R$ parameter\footnote{This parameter is the number ratio of HB
to Red Giant Branch stars brighter than the HB level} in a large
sample of Galactic globular clusters. In any case, RR Lyrae models
with $Y$=0.20 and 0.35 (Marconi, private communication) show that
reasonable variations of the helium content ($\Delta Y=\pm$ 0.03) in
metal-poor stars have quite negligible effects on the effective
temperature of the instability edges ($\Delta T_e \sim \pm$80 K).
Moreover, we wish to recall that the onset of pulsation depends also
on the efficiency of convection in the star external layers, namely
on the adopted value of the mixing length parameter $l/H_p$. Since
the effect of convection is to quench pulsation and the depth of
convection increases from high to low effective temperatures, we
expect that varying the $l/H_p$ value will modify the effective
temperature at FRE by a larger amount with respect to FBE or FOBE.
Indeed, the additional computations with $l/H_p$=2.0 computed by
Marconi \& Di Criscienzo (2007) have confirmed the general trend
shown by RR Lyrae (Marconi et al. 2003; Di Criscenzo et al. 2004)
and Classical Cepheid models (Fiorentino et al. 2007) in that, at
constant mass and luminosity, the FBE and FRE effective temperatures
increase by $\sim$ 100 and $\sim$ 300 K, respectively. The effects
of different $l/H_p$ values on the predicted relations will be
discussed in the following section.

\begin{table}[h]
\begin{center}
\caption{Chemical compositions of the adopted evolutionary models.} \label{PLCM}
\begin{tabular}{lccc}
\hline \hline
$Z$  & $Y$ & [$\alpha$/Fe] & [Fe/H]\\
\hline
0.0001 & 0.245 & 0.4 & $-$2.62\\
0.001  & 0.246 & 0.4 & $-$1.62\\
0.004  & 0.251 & 0.4 & $-$1.01\\
0.0001 & 0.245 & 0.0 & $-$2.27\\
0.001  & 0.246 & 0.0 & $-$1.27\\
0.004  & 0.251 & 0.0 & $-$0.66\\
\hline
\end{tabular}
\end{center}
\end{table}

\begin{figure}
\includegraphics[width=8cm]{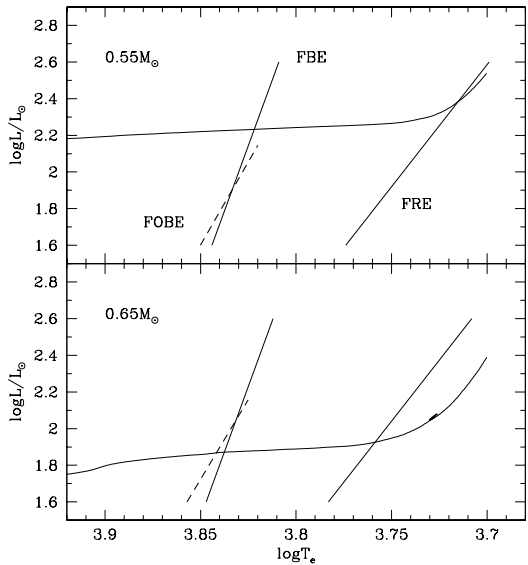}
\caption{\small{Selected evolutionary tracks with $Z$=0.0001 and
[$\alpha$/Fe]=0.4 in comparison with the predicted FOBE, FBE and
FRE. At luminosity levels brighter than the intersection between
FOBE and FBE, we found only fundamental models.}}
\end{figure}

For the evolutionary framework, we adopt the models computed by
Pietrinferni et al.~(2004, 2006) for scaled solar and
$\alpha$-enhanced ([$\alpha$/Fe]=0.4) metal distributions in order
to cover (see Table~2) the [Fe/H] range between $-2.6$ and $-0.7$.
All the models have been transferred from the theoretical HR diagram
to the various observational planes by adopting updated
color-effective temperature relations and bolometric corrections
(see Pietrinferni et al. 2004 and Cassisi et al. 2004) and the
reader is referred to these papers\footnote{The whole set of stellar
models can be retrieved at the following URL site:
http://www.te.astro.it/BASTI/index.php.} for information on the
physical inputs and numerical assumptions. Here, it seems sufficient
to note that this evolutionary framework is based on the most
updated physical scenario and that the various stellar models have
been followed all along the main core H-burning phase and advanced
core and shell He-burning evolutionary phases. All the He-burning
models adopted in present analysis have been computed by accounting
for a He-core mass and He-envelope abundance on the Zero Age
Horizontal Branch (ZAHB) characteristic on an Red Giant Branch (RGB)
progenitor with initial total mass equal to $\sim0.8M_\odot$,
corresponding to an age at the RGB tip of the order of 13~Gyr. The
reliability and accuracy of the whole evolutionary scenario have
already been tested by comparison with various empirical data sets
(see also Riello et al.~(2003); Salaris et al.~(2004); Recio-Blanco
et al. 2005) and, in summary, it appears quite suitable for
investigating the populations of variable stars in Galactic globular
clusters. Finally, let us note that these evolutionary computations
represent, so far, the
 most updated and complete set of low-mass, He-burning
 models currently available. As a fact,
the unique set of similar stellar models is
 the one published long time ago by Dorman, Rood \&
 O'Connell (1993), which is based
 on physical inputs no more updated. However, in the following section
we will discuss how the uncertainties in the evolutionary framework
 would affect the predicted relations.

\begin{table}
\caption{Luminosity $L_{FBE}$ at the fundamental blue edge of the
models with the labeled metal content and [$\alpha$/Fe]=0.4, in
comparison with the luminosity $L_{IP}$ at the intersection between
FOBE and FBE. The models in bold face are among those adopted for
deriving the predicted relations (see text). The luminosity values
are in solar units.}
\begin{center}                          % used for centering table
\begin{tabular}{lcccc}        % centered columns (4 columns)
\hline\hline                 % inserts double horizontal lines
$M/M_{\odot}$ & log$L_{FOBE}$  & log$L_{FOBE}$  & log$L_{FOBE}$  & log$L_{IP}$ \\
              & $Z$=0.0001     &  $Z$=0.001     &    $Z$=0.004   & \\
\hline
0.500 & --          & --          & 2.62       & 1.73\\
0.505 & --          & --          & {\bf 2.39} & 1.74\\
0.510 & --          & 2.55        & {\bf 2.28} & 1.74\\
0.515 & 3.08        & 2.43        & {\bf 2.20} & 1.75\\
0.520 & 2.79        & {\bf 2.32}  & {\bf 2.15} & 1.76\\
0.525 & 2.55        & {\bf 2.25}  & {\bf 2.08} & 1.77\\
0.530 & 2.46        & {\bf 2.21}  & {\bf 2.02} & 1.78\\
0.535 & {\bf 2.38}  & {\bf 2.16}  & {\bf 1.99} & 1.78\\
0.540 & {\bf 2.32}  & {\bf 2.11}  & {\bf 1.94} & 1.79\\
0.545 & {\bf 2.27}  & {\bf 2.07}  & {\bf 1.89} & 1.80\\
0.550 & {\bf 2.24}  & {\bf 2.04}  & {\bf 1.83} & 1.81\\
0.560 & {\bf 2.16}  & {\bf 1.98}  &   --       & 1.82\\
0.570 & {\bf 2.10}  & {\bf 1.95}  &   --       & 1.84\\
0.580 & {\bf 2.05}  & {\bf 1.88}  &   --       & 1.85\\
0.590 & {\bf 2.02}  & {\bf 1.87}  &   --       & 1.86\\
0.600 & {\bf 2.01}  & --          &   --       & 1.88\\
0.610 & {\bf 1.96}  & --          &   --       & 1.89\\
0.620 & {\bf 1.93}  & --          &   --       & 1.91\\
\hline
\end{tabular}
\end{center}
\end{table}

The procedure for deriving the observational parameters of the
predicted pulsators is in principle quite simple and has been
described in several previous investigations (Bono, Caputo \&
Santolamazza 1997; Marconi et al. 2003; Fiorentino et al. 2006). As
shown in Fig. 1, the relations of the predicted edges of the
instability strip [eqs. (2)-(4)] give us the way to select the
models evolving with a luminosity larger than log$L_{IP}$ and
showing FBE$\ge$ log$T_e\ge$ FRE. In this way, we derive that the
mass range of the predicted fundamental pulsators varies from
0.515-0.62\msun at [Fe/H]=$-$2.6 to 0.50-0.55\msun at [Fe/H]=$-$0.7.
These mass values are in coherence with the mass range of the
pulsating models listed in Table 1. However, as shown in Table 3,
the average luminosity of the pulsators with a given mass increases
as the metal content decreases, yielding that the [Fe/H]=$-$2.6
pulsators less massive than 0.53\msun are more luminous than our
brightest pulsating models. Since we cannot {\it a priori} be sure
that the edge and period relations provided by the pulsation models
listed in Table 1 can be extrapolated to higher luminosity levels,
in the following we will use only the predicted pulsators whose mass
and luminosity are consistent with those adopted for the pulsation
models, as given in bold face in Table 3.

\begin{figure}
\includegraphics[width=8cm]{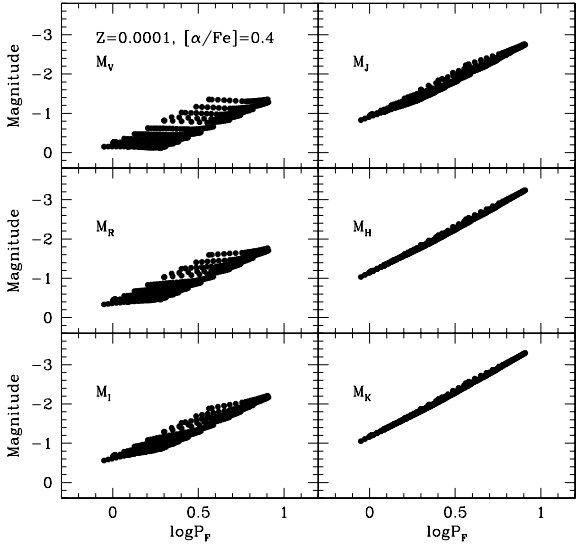}
\caption{\small{Period-Magnitude diagrams of fundamental pulsators
with $l/H_p$=1.5, $Z$=0.0001 and [$\alpha$/Fe]=0.4.}}
\end{figure}

\section{The connection between stellar evolution and pulsation}

By calculating the fundamental period by means of Eq. (5) and
adopting the magnitudes computed by Pietrinferni et al. (2004,
2006), we show in Fig. 2 selected Period-Magnitude ($PM$) diagrams
of the predicted fundamental pulsators with $l/H_p$=1.5, $Z$=0.0001
and [$\alpha$/Fe]=0.4. Note that the resulting periods are in the
range of about 0.8 to 8 days, making our theoretical investigations
quite appropriate for the analysis of observed BL Her stars.

\begin{table}[h]
\begin{center}
\caption{Predicted $PM_I$, $PM_J$, $PM_H$ and $PM_K$ relations for
fundamental pulsators with iron content in the range of
[Fe/H]=$-$2.6 to $-$0.7 and $P\le$ 8 days.} \label{PLCM}
\begin{tabular}{lcccc}
\hline \hline
\multicolumn{5}{c}{$M_i=a+b$log$P_F+c$[Fe/H]+$d(l/H_p-1.5)$}\\
\hline
$M_i$   & $a$     & $b$      & $c$ &$d$    \\
\hline
$M_I$ & $-0.26\pm0.19$ & $-2.10\pm0.06$  &  +0.04$\pm$0.01 &$-$0.24\\
$M_J$ & $-0.64\pm0.13$ & $-2.29\pm0.04$  &  +0.04$\pm$0.01 &$-$0.16\\
$M_H$ & $-0.95\pm0.06$ & $-2.34\pm0.02$  &  +0.06$\pm$0.01 &$-$0.08 \\
$M_K$ & $-0.97\pm0.06$ & $-2.38\pm0.02$  &  +0.06$\pm$0.01 &$-$0.06 \\
\hline
\end{tabular}
\end{center}
\end{table}

As already found for other pulsating variables, the effect of the
intrinsic width in effective temperature of the instability strip
(see Fig. 1) is greatly reduced when moving from optical to
near-infrared magnitudes. On this basis, it is quite clear that
synthetic $PM_B$ to $PM_R$ relations will significantly depend on
the actual distribution of the pulsators within the pulsation
region, at variance with the case of the near-infrared magnitudes.
Moreover, we wish to recall that a variation of the mixing
length parameter from $l/H_p$=1.5 to 2.0 gives hotter FBE and FRE by
about 100 K and 300 K, respectively. Consequently, the pulsator
distribution is slightly shifted toward shorter periods, yielding
mildly steepened (less than 2\%) and brighter $PM$ relations, mainly
in the optical bands. A least square fit to all the fundamental
models yields the linear relations listed in Table 4: as a result,
we get that the apparent distance modulus $\mu_I$ of observed
variables can be determined within $\pm$ 0.20 mag, including the
uncertainty due to the mixing-length parameter, whereas either
$\mu_H$ and $\mu_K$ can be determined with a formal accuracy of 0.07
mag. Concerning the $PM_J$ relation, given the residual effect of
the intrinsic width of the instability strip, it yields $\mu_J$
within 0.15 mag.

\begin{figure}
\includegraphics[width=8cm]{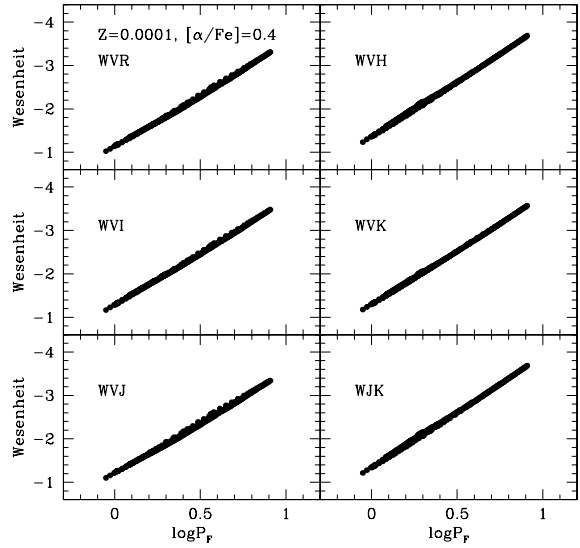}
\caption{\small{Selected Period-Wesenheit diagrams of fundamental
pulsators with $l/H_p$=1.5, $Z$=0.0001 and [$\alpha$/Fe]=0.4.}}
\end{figure}

\begin{figure}
\includegraphics[width=8cm]{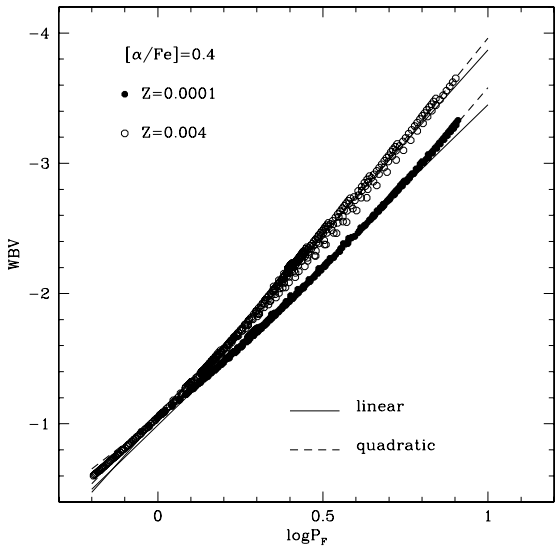}
\caption{\small{$WBV$ function versus period of selected fundamental
pulsators with $l/H_p$=1.5. The solid and dashed lines are the
linear and quadratic fit, respectively.}}
\end{figure}

It is widely acknowledged that the scatter in optical magnitudes
can be removed if a Period-Magnitude-Color ($PMC$) is considered,
i.e., if the pulsator magnitude is given as a function of the period
and color. Several previous papers (see, e.g., Madore 1982; Madore
\& Freedman 1991; Tanvir 1999; Caputo et al. 2000, 2004)
have already shown that the color coefficient of the various
$PMC$ relations is not too different from the
extinction-to-reddening ratio provided by optical and near-infrared
reddening laws (see Dean et al. 1978; Caldwell \& Coulson 1987;
Cardelli, Clayton \& Mathis 1989; Laney \& Stobie 1993). On this
basis, the adoption of the reddening insensitive Wesenheit
functions, where the magnitude is corrected for the color according
to the interstellar extinction, removes also the largest part of the
effect of differing effective temperatures. In the following,
adopting $A_V=3.1E(B-V)$, $A_R=2.45E(B-V)$, $A_I=1.85E(B-V)$,
$A_J=0.897E(B-V)$, $A_H=0.574E(B-V)$ and $A_K=0.372E(B-V)$, we will
refer to the Wesenheit functions $WBV=V-3.1E(B-V)$,
$WVR=V-4.77(V-R)$, $WVI=V-2.48(V-I)$, $WVJ=V-1.407(V-J)$,
$WVH=V-1.227(V-H)$ and $WVK=V-1.136(V-K)$. Moreover, since only
near-infrared data are available for several P2Cs, we will
consider also the function $WJK$=$K-0.709(J-K)$.

\begin{table}[h]
\begin{center}
\caption{Period-Wesenheit relations for fundamental pulsators with
iron content in the range of [Fe/H]=$-$2.6 to $-$0.7 and period
$P\le$ 8 days.} \label{PLCM}
\begin{tabular}{lcccc}
\hline
\hline
\multicolumn{5}{c}{$W=a+b$log$P_F+c$[Fe/H]+$d(l/H_p-1.5)$}\\
\hline
$W$ & $a$ &  $b$ & $c$ &$d$ \\
\hline
$WVR$ & $-1.07\pm$0.07  & $-2.42\pm$0.02  &   +0.01$\pm$0.01 &$-$0.10  \\
$WVI$ & $-1.16\pm$0.07  & $-2.43\pm$0.02  &   +0.04$\pm$0.01 &$-$0.10 \\
$WVJ$ & $-1.04\pm$0.06  & $-2.37\pm$0.02  &   +0.05$\pm$0.01 &$-$0.08 \\
$WVH$ & $-1.20\pm$0.06  & $-2.58\pm$0.02  &   +0.06$\pm$0.01 &$-$0.07 \\
$WVK$ & $-1.13\pm$0.06  & $-2.52\pm$0.02  &   +0.06$\pm$0.01 &$-$0.06\\
$WJK$ & $-1.15\pm$0.06  & $-2.60\pm$0.02  &   +0.06$\pm$0.01 &$-$0.06 \\
\hline
\end{tabular}
\end{center}
\end{table}

As shown in Fig. 3, where the fundamental pulsators with
$l/H_p$=1.5, $Z$=0.0001 and [$\alpha$/Fe]=0.4 are plotted is some
selected Period-Wesenheit diagrams, the magnitude dispersion at
fixed period is indeed greatly reduced, leading to tight linear $PW$
relations. With regard to the effect of an increased value of
the mixing length parameter, we find that moving from $l/H_p$=1.5 to
2.0 yields slightly brighter $PW$ relations, while leaving almost
unvaried (less than 2\%) the slope. By a least square  fit to all
the fundamental pulsators, we derive the coefficients listed in
Table 5. These relations give us a quite safe way to estimate the
intrinsic distance modulus $\mu_0$ of observed variables with a
formal accuracy of $\sim$ 0.1 mag, independently of the reddening.
Concerning the $WBV$ function, we show in Fig. 4 that the pulsator
distribution in the log$P_F$-$WBV$ plane is much better represented
by a quadratic relation, i.e.,
$WBV=\alpha+\beta$log$P_F+\gamma$(log$P_F)^2$, mainly at the lower
metal content. Note also that, at variance with the other Wesenheit
functions, the $WBV$ function becomes brighter as the pulsator metal
content increases, at fixed period, with the magnitude difference
increasing towards the longer periods. As a whole, the least square
fit to all the fundamental pulsators yields
$\alpha=-1.06$($\pm$0.09), $\beta=-2.96(\pm0.08)-0.36$[Fe/H] and
$\gamma=-0.17(\pm0.05)+0.13$[Fe/H].

In closure of this section, let us finally note that the
dependence of the HB luminosity at the RR Lyrae gap on the metal
content $Z$ seems to be a robust result of stellar evolution and
that all the available sets of evolutionary models, but few
exceptions, predict similar trend (see Fig. 13 in Pietrinferni et
al. 2004). Unfortunately, no comparison with other recent models can
be made for He-burning low-mass models and for this reason we adopt
an uncertainty of about $\pm$0.04 dex of the logarithm luminosity as
a safe estimate. However, when accounting for the dependence of the
pulsation period on the stellar luminosity, this uncertainty on the
stellar brightness  {\it has no significative effects on the
predicted $PM$ and $PW$ relations given in Table 4 and Table 5,
respectively.} Indeed, an increase in the luminosity by 0.04 dex,
for any fixed effective temperature, causes a period variation
$\delta$log$P$=0.036 while all the magnitudes and Wesenheit
functions become brighter by 0.1 mag. As a consequence of these
simultaneous variations, the ``new'' $PM$ and $PW$ relations will be
brighter by 0.02 mag at most.

\section{Comparison with observations}
The Galactic globular clusters with observed P2Cs are listed in
Table 6 with their reddening $E(B-V)$, apparent visual magnitude
$V$(HB) and HB type, as given by Harris (1996). We recall that the
HB type is the ratio (B$-$R)/(B+V+R), where V is the number of RR
Lyrae variables, while B and R are the numbers of HB stars bluer and
redder, respectively, than RR Lyrae stars. For all the P2Cs, we will
adopt the periods and the apparent magnitudes provided by Pritzl et
al. (2003, $BVI$, hereafter Pr03) and Matsunaga et al. (2006,
$JHK_s$, hereafter Ma06).

\begin{table*}[h]
\begin{center}
\caption{Galactic globular clusters with observed P2Cs listed with
their reddening $E(B-V)$, iron content [Fe/H], apparent visual
magnitude $V$(HB) and HB type, as given by Harris (1996). The last
two columns give the numbers of BL Her and W Vir stars.}
\label{PLCM}
\begin{tabular}{lcccccc}
\hline\hline
Name    &   $E(B-V)$ &  [Fe/H] &    $V$(HB) &   HB type &     N$_{BL}$ & N$_{WV}$\\
\hline
HP1 &   1.19    &   $-$1.55 &   18.60   &    -- &   0   &   2 \\
N1904   &   0.01    &   $-$1.57 &   16.15   &   +0.89   &   0   &   1 \\
N2419   &   0.11    &   $-$2.12 &   20.45   &   +0.86   &   1   &   0   \\
N2808   &   0.22    &   $-$1.15 &   16.22   &   $-$0.49 &   1   &   0   \\
N4372   &   0.39    &   $-$2.09 &   15.50   &   +1.00   &   2   &   0   \\
N5139-$\omega$ Cen  & 0.12 & $-$1.62    &   14.53   &   +0.94   &   9   &   2   \\
N5272-M3    &   0.01    &   $-$1.57 &   15.68   &   +0.08   &   0   &   1   \\
N5904-M5    &   0.03    &   $-$1.27 &   15.07   &   +0.31   &   0   &   2   \\
N5986   &   0.28    &   $-$1.58 &   16.52   &   +0.97   &   0   &   1   \\
N6093-M80   &   0.18    &   $-$1.75 &   16.10   &   +0.93   &   0   &   1   \\
N6205-M13   &   0.02    &   $-$1.54 &   15.05   &   +0.97   &   4   &   1   \\
N6218-M12   &   0.19    &   $-$1.48 &   14.60   &   +0.97   &   0   &   1   \\
N6229   &   0.01    &   $-$1.43 &   18.03   &   +0.24   &   0   &   1   \\
N6254-M10   &   0.28    &   $-$1.52 &   14.65   &   +0.98   &   1   &   2   \\
N6256   &   1.03    &   $-$0.70  &   18.50   &    -- &   0   &   1   \\
N6266-M62   &   0.47    &   $-$1.29 &   16.25   &   +0.32   &   0   &   1   \\
N6273-M19   &   0.41    &   $-$1.68 &   16.50   &    -- &   1   &   2   \\
N6284   &   0.28    &   $-$1.32 &   17.40   &    -- &   2   &   0   \\
N6293   &   0.41    &   $-$1.92 &   16.50   &   +0.90   &   1   &   0   \\
N6325   &   0.89    &   $-$1.17 &   17.90   &    -- &   0   &   2   \\
N6341-M92   &   0.02    &   $-$2.28 &   15.10   &   +0.91   &   1   &   0   \\
N6388   &   0.37    &   $-$0.60  &   16.85   &    -- &   2   &   1   \\
N6402-M14   &   0.60    &   $-$1.39 &   17.30   &   +0.65   &   2   &   3   \\
N6441   &   0.47    &   $-$0.53 &   17.51   &    -- &   2   &   5   \\
N6453   &   0.66    &   $-$1.53 &   17.53   &    -- &   0   &   2   \\
N6569   &   0.55    &   $-$0.86 &   17.52   &    -- &   0   &   1   \\
N6626-M28   &   0.40    &   $-$1.45 &   15.55   &   +0.90   &   0   &   2   \\
N6626-M28   &   0.40    &   $-$1.45 &   15.55   &   +0.90   &   0   &   1   \\
N6715-M54   &   0.15    &   $-$1.58 &   18.17   &   +0.75   &   2   &   0   \\
N6749   &   1.50    &   $-$1.60  &   19.70   &   +1.00   &   1   &   0   \\
N6752   &   0.04    &   $-$1.56 &   13.70   &   +1.00   &   1   &   0   \\
N6779-M56   &   0.20    &   $-$1.94 &   16.16   &   +1.00   &   1   &   1   \\
N7078-M15   &   0.10    &   $-$2.26 &   15.83   &   +0.67   &   2   &   1   \\
N7089-M2    &   0.06    &   $-$1.62 &   16.05   &   +0.96   &   0   &   4   \\
Ton1    &   2.28    &   $-$1.30  &   21.40   &    -- &   0   &   1   \\
\hline
\end{tabular}
\end{center}
\end{table*}

\begin{figure}
\includegraphics[width=8cm]{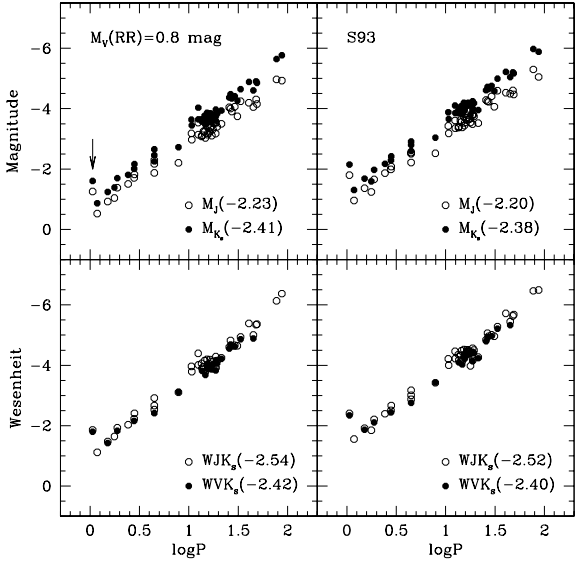}
\caption{\small{$PM$ and $PW$ distributions of observed P2Cs under
two different assumptions on the absolute magnitude of RR Lyrae
stars. The numbers in parentheses are the slopes of the relations,
as derived by a linear regression to the data. The arrow indicates
V7 in NGC 6341. The infrared magnitudes are taken by Matsunaga et
al. (2006). }}
\end{figure}

\begin{figure}
\includegraphics[width=8cm]{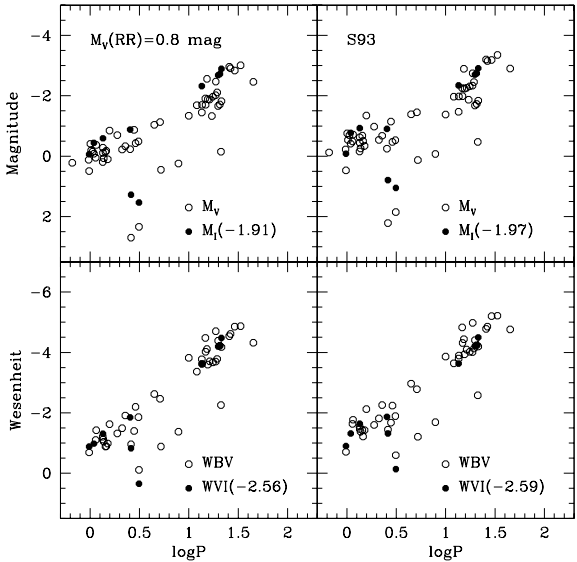}
\caption{\small{$PM$ and $PW$ distributions of observed P2Cs under
two different assumptions on the absolute magnitude of RR Lyrae
stars. The outliers are V24 and V28 in NGC 4372, V12 and V32 in NGC
6205 and V3 in NGC 6254. The optical magnitudes are taken by Pritzl
et al. (2003).}}
\end{figure}

In their investigations, Pr03 and Ma06  find quite tight linear
correlations of the absolute magnitudes of P2Cs, as derived from
RR Lyrae based distance moduli, with log$P$, without a clear
evidence for a change in the slope between BL Her and W Vir stars.
In particular, adopting $M_V$(RR)=0.89+0.22[Fe/H], Ma06 derive
$\delta M_J/\delta$log$P=-2.23(\pm$0.05), $\delta
M_H/\delta$log$P=-2.34(\pm$0.05) and $\delta
M_{K_s}/\delta$log$P=-2.41(\pm$0.05), which are in
excellent agreement with the
predicted slopes given in Table 4. In that study, it is also
mentioned that the slope of the observed near-infrared $PM$
relations is not affected by different assumptions on the slope of the
$M_V$(RR)-[Fe/H] relation. As a fact, by repeating the Ma06
procedure but adopting $M_V$(RR)=0.8 mag and
$M_V$(RR)=0.94+0.30[Fe/H] (Sandage 1993: S93), we show in Fig. 5
(upper panels) that the
variation to the near-infrared $PM$ slopes is less
than 2 percent. Here, we add that this holds also for the $PW$
relations based on $VJHK_s$ magnitudes (lower panels). Note that the
arrow in this figure refers to NGC 6341 V7 (Del Principe et al.
2005) which will be discussed separately.

Turning to the Pr03 optical magnitudes, some comments should be made
to the data plotted in Fig. 6. Firstly, even removing the too faint
outliers (i.e., the $BVI$ data of V24 and V28 in NGC 4372 and the
$BV$ data of V12 and V32 in NGC 6205 and of V3 in NGC 6254, see also
Pr03), we note a significant dispersion in the $PM_V$ and $PWBV$
planes as well as some evidence that the W Vir stars follow steeper
$PM_V$ relations than the BL Her stars. As for the observed $PM_I$
and $PWVI$ relations, they appear linear and quite tight, with the
slope independent of the adopted $M_V$(RR)-[Fe/H] relation and in good
agreement with the predicted values. However,
it should be mentioned that these results are based on a rather
small number of data points (mostly, the variables in NGC 6441).

In summary, we derive that all the observed P2Cs show linear
near-infrared $PM$ relations and linear $PW$ relations, with the
exclusion of $WBV$, independently of the adopted $M_V$(RR)-[Fe/H]
relation. Moreover, the slopes of these relations are in close
agreement with our predicted values for variables with $P\le$ 8
days, supporting the hypothesis of similar relations for BL Her and
W Vir stars. In addition, we wish to mention that the [Fe/H] effect
on the zero-point of the near-infrared $PM$ relations, as estimated
by Ma06, is about 0.1 mag/dex$^{-1}$, again in agreement with our
theoretical value.

On this ground, we can use the predicted relations derived in the
previous section to derive the P2C distance moduli. With this
purpose, since the Pietrinferni et al. (2004, 2006) magnitudes are
in the Bessell \& Brett (1988) near-infrared photometric system, the
relations provided by Carpenter (2001) are used to transform the
original 2MASS $JHK_s$ data given by Ma06 into standard $JHK$
magnitudes. For the P2C reddening and metal content, we adopt the
values of the hosting globular cluster; however, for NGC 6388 and
NGC 6441 we consider also [Fe/H]=$-$2.0, as adopted by Pr03 and
Ma06.

By excluding the globular clusters with only W Vir stars and
NGC 6341 V7, which will be discussed separately, we find some points
worthy of mention:

\begin{itemize}
\item   from the $BVI$ magnitudes of the NGC 6441 P2Cs, we find an increasing
discrepancy between the $\mu_{0,WBV}$ and $\mu_{0,WVI}$ values of a
given variable when moving from short to long period stars. With
[Fe/H]=$-$0.53, we get $\mu_{0,WBV}-\mu_{0,WVI}\sim$ 0.4 mag at
log$P\sim$ 1.1 and $\sim$ 0.8 mag at log$P\sim$ 1.3. It is
interesting to note that, if [Fe/H]=$-$2.0 is adopted, these
differences are reduced to $\sim$ 0 and $\sim$0.3 mag, respectively.
 This would suggest that, despite the cluster high metallicity, the
P2Cs may have a low metal content. However, we remind that the
$\mu_{0,WBV}$ values are expected to be
affected by large
errors as a result of the older and less accurate photometry in these bands;

\item   for each given variable, the intrinsic distance modulus inferred by the
near-infrared $WJK$ function is in a general agreement with the
values based on the $WVJ$, $WVH$ and $WVK$ functions. No sure
comparison can unfortunately be made with $\mu_{0,WVI}$, whereas we
note significant discrepancies with the results based on $WBV$. This
can be due to some old $BV$ data (see also Pr03), as well as, in the
case of  $\omega$ Cen variables, to the occurrence of a
metallicity spread;

\item       except the variables in $\omega$ Cen,
the trend of the intrinsic and the apparent distance moduli, as
determined by $JHK$ magnitudes, yields $A_J/E(B-V)\sim$ 0.85,
$A_H/E(B-V)\sim$ 0.54 and $A_K/E(B-V)\sim$ 0.40, in reasonable
agreement with the extinction laws mentioned in Section 3.
\end{itemize}

\begin{table}[h]
\begin{center}
\caption{Mean intrinsic distance moduli (in magnitudes), as derived
by the Wesenheit functions of BL Her and W Vir stars. The quantity N
is the number of averaged values. As a matter of comparison, in the
last column we give the intrinsic distance moduli based on RR Lyrae
stars (see text).} \label{PLCM}
\begin{tabular}{lccccc}
\hline\hline
name    &   $\mu_0$(BLH)      & N  & $\mu_0$(WV) & N &$\mu_0$(RR)    \\
\hline
N2808   &   14.99$\pm$0.06    & 1  &  --             &  --& 14.90\\
N5139   &   13.84$\pm$0.08    & 4  &  13.78$\pm$0.16 & 4 & 13.62  \\
N6254   &   13.39$\pm$0.20    & 4  &  13.56$\pm$0.16 & 2 &13.23  \\
N6273   &   14.69$\pm$0.06    & 1  &  14.77$\pm$0.06 & 2 &14.71  \\
N6284   &   15.77$\pm$0.06    & 2  &   --            & --&15.93    \\
N6293   &   14.80$\pm$0.06    & 1  &   --            &  -- &14.76\\
N6402   &   14.80$\pm$0.08    & 8  &  14.93$\pm$0.09 & 8 &14.86  \\
N6441   &   15.55$\pm$0.07    & 2  &  15.53$\pm$0.19 & 6 & 15.28  \\
N6441mp &   15.61$\pm$0.07    & 2  &  15.58$\pm$0.18 & 6 & 15.60   \\
N6715   &   17.19$\pm$0.10    & 2  &  --             & --& 17.16\\
N6749   &   14.56$\pm$0.08    & 1  &  --             &  --&14.51 \\
N6779   &   15.05$\pm$0.06    & 4  &  15.31$\pm$0.11 &4  &15.08 \\
\hline
\end{tabular}
\end{center}
\end{table}

Eventually, by excluding the $\mu_{0,WBV}$ values, we derive the
mean intrinsic distance moduli listed in Table 7 together with the
estimated total uncertainty. As a matter of comparison, we give in
the last column the cluster distance determined  by adopting the
relation $M_V$(RR)=0.89+0.22[Fe/H] and the reddening and HB visual
magnitude in Table 6. As a whole, given the well known debate on the
RR Lyrae distance scale (see the recent review by Cacciari \&
Clementini 2003), we believe that the P2Cs distances agree with the
RR Lyrae based values within the estimated uncertainty. As for a
change in the $PM$ and $PW$ slopes around log$P \sim$ 1, namely
between BL Her and W Vir stars, the values listed in Table 8 show
similar distances for the variables in $\omega$ Cen and NGC 6441,
whereas for the remaining clusters there is a subtle discrepancy as
the former distance moduli are shorter by $\sim$ 0.15 mag than the
latter ones.

We can now study NGC 6341 V7 which has a period log$P$=0.026 and is
deviant from the near-infrared $PM$ relations (see arrow in Fig. 5).
With $VJHK$ data taken from Del Principe et al. (2005), we would
derive $\mu_0$=13.80$\pm$0.07 mag or 14.10$\pm$0.07 mag, depending
on whether the variable is a fundamental or first-overtone BL Her
star (i.e, adopting log$P_F$=0.038). As a whole, these distances
are too short
with respect to $\mu_0(RR)$=14.65 mag, as derived by the relation
$M_V$(RR)=0.89+0.22[Fe/H] using the values listed in Table 6.  On
the other hand, comparing the $K$ magnitudes of RR Lyrae stars from
Del Principe et al. (2005) with the predicted $PM_K$ relations
presented by Del Principe et al. (2006), we derive
$\mu_0(RR)$=14.63$\pm$0.08 mag. Since the relations given in the
present paper and those reported by Del Principe et al. (2006) are
based on model computations which adopt homogeneous physics and
numerical procedures, we can conclude that
V7 is {\it not} a BL Her star but, due to the evidence that it is
brighter than expected for its period, it may be an Anomalous
Cepheid. Indeed, using for V7 the $PM_K$ relations determined by Fiorentino
et al. (2006) from evolutionary and pulsation models of fundamental
ACs, we derive $\mu_K$(AC)=14.69$\pm$0.15 mag.

\section{Conclusions}
The main results of this study can be summarized as follows:
\begin{itemize}
\item   On the basis of updated nonlinear convective pulsation models of
BL Herculis we derive analytical relations for the boundaries
of the instability strip as a function of the adopted stellar
physical parameters, as well as the pulsation equation for these
bright objects. Moreover, we confirm earlier suggestions that for
each given mass and helium content there exists an ``intersection''
luminosity, as given by the intersection between the FOBE and the
FBE, above which only the fundamental mode is stable.
\item   By combining the pulsational results with the predictions of the
evolutionary models by Pietrinferni et al. (2004, 2006), we select
models brighter than the ``intersection'' luminosity and
therefore pulsating only in the fundamental mode, corresponding to
periods longer than $\sim$ 0.8d and stellar masses $\le 0.62
M_{\odot}$.
\item For the models which have masses and luminosities consistent
with the physical parameters adopted in the pulsation models, we
derive the predicted $PM$ and $PW$ relations at the various
photometric bands. We show that the predicted slopes are
in close agreement with the empirical ones, quite
independently of the slope in the adopted $M_V$(RR)-[Fe/H] relation.
\item   The predicted $PM$ and $PW$ relations have been applied to all the known P2Cs
in Galactic globular clusters and the resulting distance moduli are
in statistical agreement with the RR Lyrae based values.
\item   The variables in $\omega$ Cen and NGC 6441 seem to support
the hypothesis of unique $PM$ and $PW$ relations for BL Her and W
Vir stars. Conversely, for the remaining clusters the former
distance moduli are found to be shorter by $\sim$ 0.15 mag than the
latter ones, suggesting steeper relations with $P\ge$ 10 days.
On this ground, no firm conclusion can presently be found in favor or against
a change in the $PM$ and $PW$ slopes around log$P \sim$ 1.
\item   Finally, the application of the predicted relations
to NGC6341 V7 provides evidence that this variable is not a P2C.
Using the results presented by Fiorentino et al. (2006), we confirm
the earlier suggestion by Ma06 that this star can be the second AC
in Galactic globular clusters.
\end{itemize}

\begin{acknowledgements}
Financial support
for this study was provided by MIUR, under the scientific project
``On the evolution of stellar systems:
fundamental step toward the scientific exploitation of VST''
(P.I. Massimo Capaccioli) and by INAF, under the scientific project
``A laboratory fo the theoretical study of stellar populations'' (P.I. A. Buzzoni).
\end{acknowledgements}

\pagebreak
%\begin{references}

%\end{references}

\clearpage

\end{document}